\documentclass[12pt,preprint]{aastex}

\newcommand\nd{\nodata}
\newcommand\mcnd{\multicolumn{2}{c}{\nodata}}

\newcommand{\hecs}{\sc{helio10}\upshape} 
\newcommand{\tc}{$t^2$}

\newcommand{\Hiir}{H$\,${\small\rmfamily{II}}\relax~region}
\newcommand{\Hiirs}{H$\,${\small\rmfamily{II}}\relax~regions}

\shorttitle{2 SMC \Hiirs~ Considering \tc$>$0}
\shortauthors{Pe\~na-Guerrero, Peimbert, \& Peimbert}

\received{}
\begin{document}

\title{Analysis of two SMC \ion{H}{2} Regions Considering Thermal Inhomogeneities: Implications for the Determinations of Extragalactic Chemical Abundances
\footnotemark[1]}

\footnotetext[1]{Based on observations collected at the European Southern
Observatory, Chile, proposal number ESO 69.C-0203(A)}

\author{Mar\'ia A. Pe\~na-Guerrero\footnotemark[2]}
\email{guerrero@astroscu.unam.mx}

\author {Antonio Peimbert\footnotemark[2]}
\email{antonio@astroscu.unam.mx}

\author{Manuel Peimbert\footnotemark[2]}
\email{peimbert@astroscu.unam.mx}

\and

\author{Mar\'{\i}a Teresa Ruiz\footnotemark[3]}
\email{mtruiz@das.uchile.cl}

\footnotetext[2]{Instituto de Astronom\'\i a, Universidad Nacional Aut\'onoma de M\'exico, Apdo. Postal 70-264, M\'exico 04510 D.F., M\'exico}
\footnotetext[3]{Departamento de Astronom\'ia, Universidad de Chile, Casilla Postal 36D, Santiago de Chile, Chile}

\begin{abstract}
We present long slit spectrophotometry considering the presence of thermal inhomogeneities (\tc) of two \Hiirs~in the Small Magellanic Cloud (SMC): NGC 456 and NGC 460. Physical conditions and chemical abundances were determined for three positions in NGC 456 and one position in NGC 460, first under the assumption of uniform temperature and then allowing for the possibility of thermal inhomogeneities. We determined \tc~values based on three different methods: i) by comparing the temperature derived using oxygen forbidden lines with the temperature derived using helium recombination lines, ii) by comparing the abundances derived from oxygen forbidden lines with those derived from oxygen recombination lines, and iii) by comparing the abundances derived from ultraviolet carbon forbidden lines with those derived from optical carbon recombination lines. The first two methods averaged \tc=$0.067\pm0.013$ for NGC 456 and \tc=$0.036\pm0.027$ for NGC 460. These values of \tc~imply that when gaseous abundances are determined with collisionally excited lines they are underestimated by a factor of nearly 2. From these objects and others in the literature, we find that in order to account for thermal inhomogeneities and dust depletion, the O/H ratio in low metallicity \Hiirs~should be corrected by 0.25$\,$-$\,$0.45 dex depending on the thermal structure of the nebula, or by 0.35 dex if such information is not available.

\end {abstract}


\section{Introduction}

The Magellanic Clouds play a fundamental role in astrophysics because they are the closest laboratory we have to probe the extragalactic distance scale and the theories about chemical evolution of stars and galaxies. They are close enough to be able to identify individual objects and isolate them from their environment. In the case of chemical abundance studies of \Hiirs, it is possible to perform observations that avoid bright stars, Wolf-Rayet stars, and supernova remnants; this can translate in extremely good equivalent widths in emission, EW$_{em}$(H$\beta$)$\,\gtrsim\,$250 \AA.

There have been many studies of the chemical composition of \Hiirs~in the Magellanic Clouds \citep[e.g.][]{pei76,pag78,wes90,cap96,pei00}. It has been found that the chemical composition within each cloud is uniform, and that the SMC has lower metallicity (approximately 0.40 dex) than the Large Magellanic Cloud (LMC); making the SMC the best object to study low metallicity environments. In particular, \Hiirs~of the SMC provide the means to study low metallicity star-forming neighborhoods and NGC 456 is one of the brightest \ion{H}{2} regions of the SMC \citep{pei76}. In this study the number of lines presented for NGC 456 and NGC 460 is considerably larger than in previous works.

Almost all studies of \Hiirs~use the direct method to determine gaseous abundances, and implicitly assume that this value represents the chemical abundances of the ISM. The direct method uses collisionally excited lines (CELs) to obtain chemical abundances, and presuppose a homogeneous temperature structure. When abundances are calculated using recombination lines (RLs), they are approximately factors of 2 higher than abundances obtained with CELs; this is now known as the abundance discrepancy factor (ADF) problem. Moreover, the depletion into dust grains can account for a large fraction of the oxygen atoms.

\citet{est98} measured the depletion into dust in the Orion nebula and suggested that a 0.08 dex correction to the O/H ratio should be included when considering \Hiirs. More recent studies of the Orion nebula suggest that the correction is higher \citep{mes09b,sim11} while recent studies of depletion of Mg, Si, and Fe suggest that the correction depends on the O/H ratio of the specific \Hiir~\citep{pea10}, reaching a fraction of up to 28\%~for high metallicity regions.

\citet{pei67} proposed to take into account \tc~when using CELs to determine abundances in modeling \Hiirs, and found that spatial thermal fluctuations of approximately 20\%~are capable of producing errors of factors of 2 to 3 on the determination of chemical abundances. The additional complications to calculate abundances using \tc, along with the fact that there is no tailored physical model to fit temperature inhomogeneities, have made its reception within the community poor.

Since the idea of Peimbert's temperature inhomogeneities was introduced, several explanations about where such temperature inhomogeneities could be coming from have been proposed: shock waves, shadowed regions, advancing ionization fronts, multiple ionizing sources, X-rays, magnetic reconnection, and inhomogeneous chemical composition, among others \citep[e.g.][and references therein]{pei06,tsa05}. However, most of these explanations are capable of either producing only a fraction the \tc~value found in observations or, under fine-tuned circumstances, the total observed \tc~values.

For most objects, two explanations are capable of producing the observed \tc~values: a) the observed \tc~values come from a variety of mechanisms in a chemically homogeneous medium, where each mechanism is only responsible for a fraction of the total; and b) the presence of chemical inhomogeneities within the photoionized region. As of now, no complete self-consistent model has been proposed for either explanation: models for explanation (a) have not been able to explain where all the required energy to keep these thermal inhomogeneities comes from, while models for explanation (b) have not been able to describe the exact characteristics and evolution of the proposed high metallicity inclusions. As for abundance determinations: option (a) would indicate that RLs give the correct abundances, or that one could use CELs and the formalism of \tc~of \citet{pei69} to determine abundances; and models of option (b) imply that the chemical abundance is intermediate between those obtained from CELs adopting \tc=0.00, and RLs adopting a chemically inhomogeneous medium \citep{tsa05}. For either option, the abundances determined using CELs with \tc=0.00 need a systematic correction.

Additional evidence supporting chemically homogeneous models with large \tc~values implied by the ADF problem comes from: the theoretical vs. observational $\Delta Y/\Delta O$ ratio, abundances of F and G young stars of the solar vicinity vs. abundances of \Hiirs, oxygen abundances of O and B stars vs. abundances of \Hiirs, protosolar abundances vs. \Hiir~abundances, considering the effect of Galactic chemical evolution \citep[see][and references therein]{pei11}.

In Sections \ref{obs}~and \ref{Icorr}~the observations and reduction procedure are described. In Section \ref{phys}~temperatures and densities are derived from four and three different line intensity ratios, respectively; also in this section we derive the mean square temperature inhomogeneities, $t^2$, in three different ways: i) from the comparison of the abundances of \ion{O}{2} lines to those of [\ion{O}{3}] lines, ii) using \ion{He}{1} lines, and iii) from the comparison of the abundance of \ion{C}{2} lines to those of [\ion{C}{3}] $\lambda$ 1907 + \ion{C}{3}] $\lambda$ 1909 lines. In Section \ref{IonaAb}~we determine ionic abundances with three different methods: i) based on RLs, ii) based on CELs, and iii) correcting CELs for \tc; while the total abundances derived from these methods are presented in Section \ref{totAbs}. In Section \ref{clasMR}~we present the direct method and its shortcomings, the typical correction to the O/H ratio for low metallicity \Hiirs, and the implications for the determinations of extragalactic chemical abundances. Finally, the discussion and conclusions are presented in Sections \ref{disc}~and \ref{conc}, respectively.

\section{Observations}\label{obs}

The observations were obtained during the night of September 10$^{\rm th}$ of 2002 with the Focal Reducer Low Dispersion Spectrograph 1, FORS1, at the Very Large Telescope, VLT, Melipal in Cerro Paranal, Chile. We used three grism settings: GRIS-600B+12, GRIS-600R+14 with filter GG435, and GRIS-300V with filter GG375 (see Table 1).

The slit was oriented almost east-west (position angle 82.5$^{\rm o}$) to observe the brightest regions of NGC 456 ($\alpha$=$1^h13^m50.8^s$, $\delta$=-73$^{\rm o}$18'03.2") which we called NGC 456 Position 2. The linear atmospheric dispersion corrector, LACD, was used to keep the same observed region within the slit regardless of the air mass value. The slit length was 410" and the slit width was set to 0.51", this setting was chosen to have the resolution to deblend the [\ion{O}{2}] line $\lambda$ 3726 from $\lambda$ 3729 when using the GRIS-600B+12. Three observations were made with each grism configuration to be able to identify and remove cosmic rays.

\begin{figure}
\includegraphics[angle=0,scale=0.6]{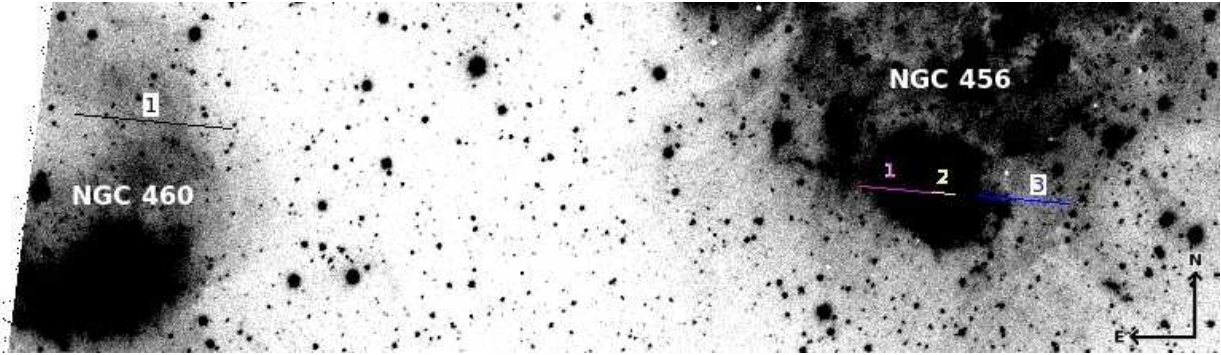}\\
(a)\\
\\
\includegraphics[angle=0,scale=0.6]{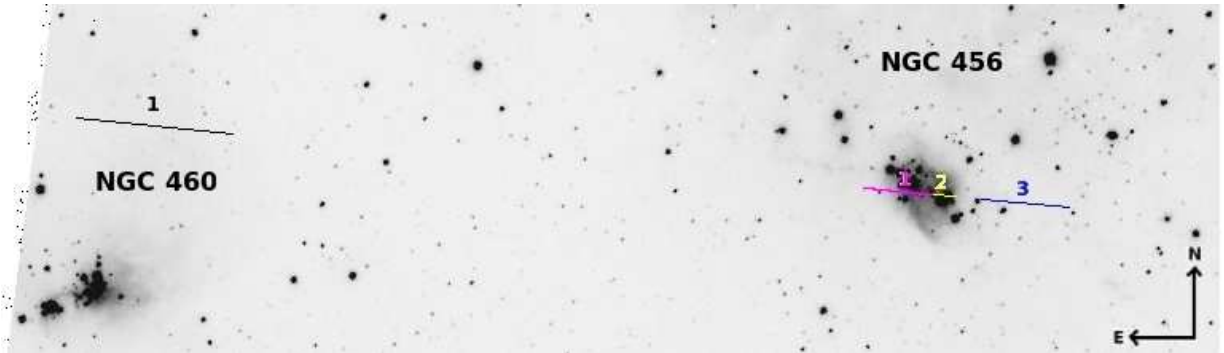}\\
(b)\\
\caption[f1.eps]{
(1a) VLT image of NGC 460 Position 1 and Positions 1, 2, and 3 of NGC 456. This image of 306'' $\times$ 87'' has Position 2 located at $\alpha=1^h13^m50.8^s$ and $\delta=-73^o18'03.2''$ (J2000.0), and was taken with a filter that suppresses light bluer than 4350 \AA. (1b) Same image but with a different saturation level so that Position 2 in NGC 456 is enhanced.\label{ngc456}}
\end{figure}

The aperture extractions were made for a length of 5.6" for NGC 456 Position 2; NGC 456 Position 1 is centered 11.4" east of Position 2 and has a length of 17.2", and NGC 456 Position 3 is centered 19.4" west of Position 2 and has a length of 22.4" (note there is a 3.2" gap to avoid a bright star between positions 2 and 3, see Figure \ref{ngc456}). The eastern part of the slit passed through the outer parts of NGC 460; we defined NGC 460 Position 1 as the region centered 199" east of NGC 456 Position 2, with a length of 40".

\begin{figure}
\begin{center}
\includegraphics[angle=0,scale=1.2]{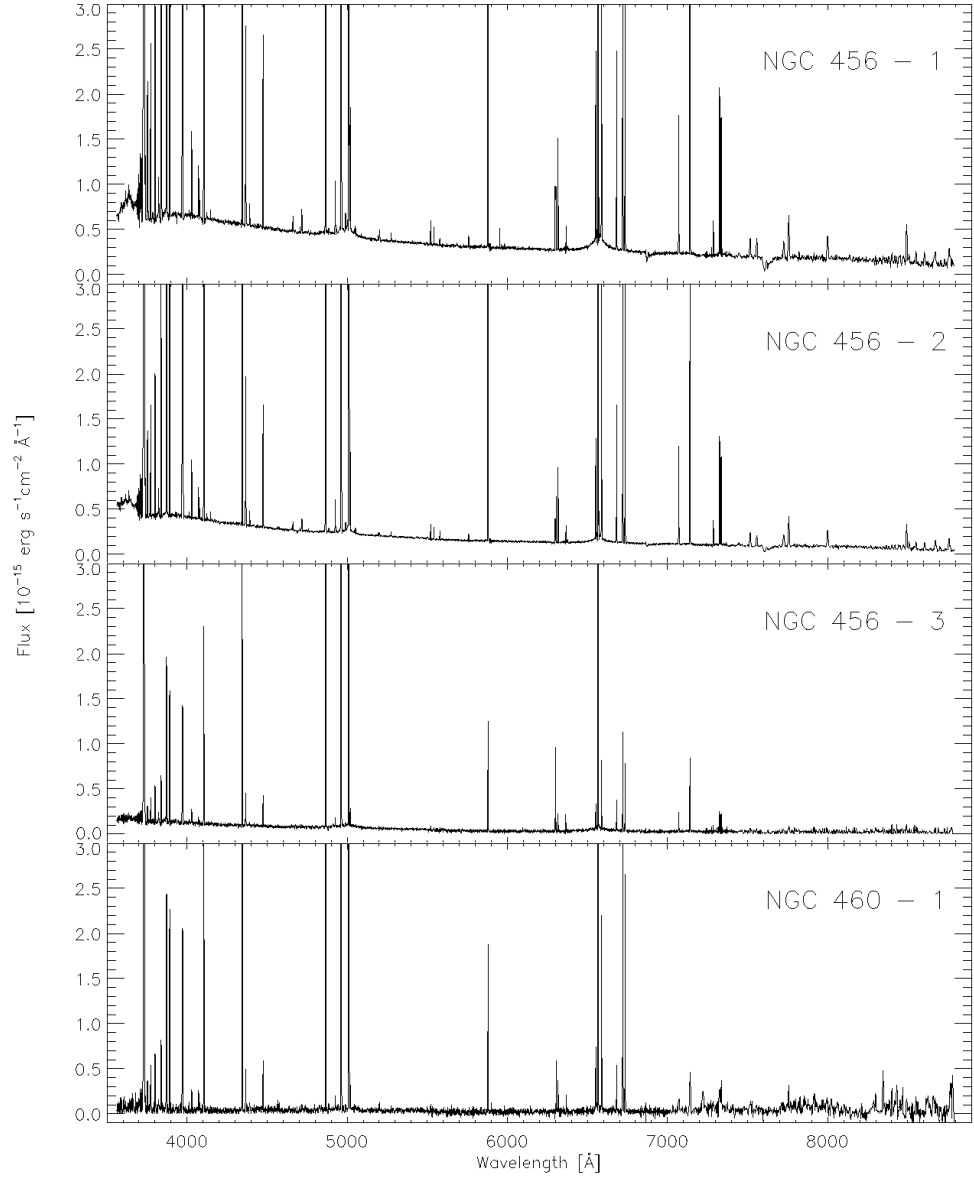}
\caption[f2.eps]{
Spectra of Positions 1, 2, and 3 in NGC 456, and Position 1 of NGC 460. These spectra include the blue (3600-5500 \AA) and red (5500-7400\AA) high-resolution spectra, as well as the low-resolution spectrum (7400-8800 \AA).\label{especs}}
\end{center}
\end{figure}

\begin{figure}
\includegraphics[angle=0,scale=0.65]{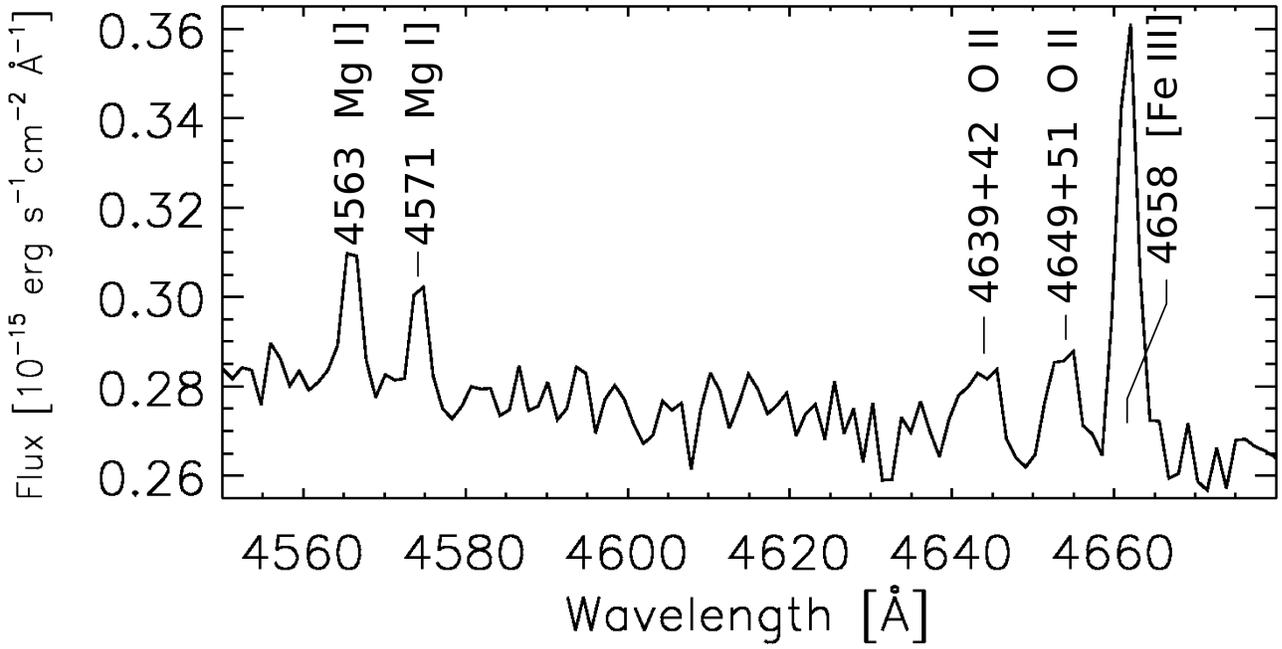}
\caption[f3.eps]{
Zoom-in of the region between $\lambda$4550 and $\lambda$4680 in NCG 456 Position 2 to show the quality of the data.\label{zoomOII}}
\end{figure}

The spectra were reduced using IRAF\footnotemark{} reduction packages, following the standard procedure of bias subtraction, aperture extraction, flat fielding, wavelength calibration and flux calibration. For flux calibration the standard stars LTT~2415, LTT~7389, LTT~7987 and EG~21 were used \citep{ham92,ham94}. The observed spectra are presented in Figure \ref{especs}; a zoom-in of the region around $\lambda$4600 of the spectrum of NGC 456-2 is presented in Figure \ref{zoomOII} to show the quality of the data.

\footnotetext{IRAF is distributed by NOAO, which is operated by AURA, under cooperative agreement with NSF.}

\section{Line Intensities and reddening correction}\label{Icorr}

The procedure of line measurement was done using the \begin{tt}splot\end{tt}~task of the IRAF package. Line intensities were measured by integrating the flux in the line between two limits over the local continuum estimated by eye. There were a few cases where line blending was found. In this situation we used a multiple Gaussian profile procedure. The error in the flux calibration has been estimated to be 1\%; the contribution to the errors due to the noise was estimated from the continuum. The final adopted errors were estimated using standard error procedures.

The reddening correction, $C$(H$\beta$), and the underlying absorption, $EW_{abs}$(H$\beta$), were fitted simultaneously to the theoretical ratios. We adopted the extinction law of \citet{sea79}, and the underlying absorption ratios obtained from the work by \citet{gon99} and from the theoretical ratios determined with the program \begin{tt}INTRAT\end{tt}~by \citet{sto95}.

The Balmer and helium emission lines were corrected for underlying absorption, and the equivalent widths in absorption adopted are presented in Table \ref{tEWs}. The extinction law values used are presented in Tables \ref{tlines}~and \ref{tlines2}. The physical conditions used for \begin{tt}INTRAT\end{tt}~were $T_e$=10,000 K and $n_e$=100 cm$^{-3}$, this is only a first guess, but since the hydrogen lines are nearly independent from temperature and density no further corrections were necessary.

The emission line intensities of all positions in NGC 456 and NGC 460 are presented in Tables \ref{tlines}~and \ref{tlines2}, respectively. Table \ref{tlines}, column (1) presents the adopted laboratory wavelength, $\lambda$, column (2) the identification for each line, column (3) presents the extinction law value used for each line \citep{sea79}, $f$($\lambda$). Columns (4$\,$-$\,$6) present the data from NGC 456-1, showing: the observed flux relative to H$\beta$, $F(\lambda)$; the flux corrected for reddening relative to H$\beta$, $I$($\lambda$); and the percentage error associated with those intensities, respectively. Columns (7$\,$-$\,$9) and (10$\,$-$\,$12) present the same data as columns (4$\,$-$\,$6) but for positions 2 and 3, respectively. In Table \ref{tlines2}~the columns are the same as the first 6 columns of Table \ref{tlines} but for Position 1 of NGC 460. The $EW_{abs}$(H$\beta$), $C$(H$\beta$), $EW$(H$\beta$), as well as the $F$(H$\beta$) and $I$($\beta$) for each position in both objects are also presented in these tables.

\section{Physical Conditions}\label{phys}

\subsection{Temperatures and densities}
The sources for the atomic data of CELs in the used IRAF are presented in Table \ref{atomic} -- for studies of nebulae with higher densities, we recommend to use more recent O$^+$ collisional strengths such as those from \citet{wan04,mon06,pra06}, and \citet{tay07}. The temperatures and densities in Table \ref{tdat} were determined using the line intensities presented in Tables \ref{tlines}~and \ref{tlines2}. These determinations were done with the \begin{tt}temden\end{tt} task in IRAF, which models populations in five-, six-, or eight-level ions to derive the physical conditions.

\subsection{Temperature Inhomogeneities}\label{Tinhom}
Instead of assuming a homogeneous temperature throughout the objects we took into account inhomogeneities in the temperature structure, which are described by the formalism developed by \citet{pei67}. To derive the ionic abundance ratios we used the average temperature, $T_0$, and the mean square temperature inhomogeneities, $t^2$, defined as follows:

\begin{equation} T_0(ion) = \frac{\int T_e(\textbf{r})N_e(\textbf{r})N_{ion}(\textbf{r}) dV}{\int N_e(\textbf{r})N_{ion}(\textbf{r}) dV}
\label{T0}, \end{equation}

\begin{equation} t^2(ion) \equiv \frac{\int (T_e-T_0)^2 N_e(\textbf{r})N_{ion}(\textbf{r}) dV}{T_0^2 \int N_e(\textbf{r})N_{ion}(\textbf{r}) dV},
\label{t2} \end{equation}
where $N_e$ and $N_{ion}$ are the electron and the ion densities, respectively, of the line of sight, and $V$ is the observed volume.

Two independent temperatures are required to infer $T_0$ and $t^2$: one that weights preferentially the high-temperature regions and another that weights preferentially the low-temperature regions \citep{pei67}. The temperatures that weight preferentially the high temperature region were defined as in \citet{pea02}:

\begin{equation} T_{4363/5007} = T_{0} \Bigg[ \bigg(1 + \frac{t^2}{2} \left( \frac{91300}{T_{0}} -3\right) \Bigg],
\label{TO3}
\end{equation}
for the O$^{++}$ region; for the O$^{+}$ region we used a similar equation for the [\ion{O}{2}] $\lambda\lambda$ 3727 and 7325 lines with an energy corresponding to 97800 K.

Following the literature \citep[][and references therein]{pea03,pea05}, we can determine a \tc~value from the helium lines. The \hecs~program determines a a temperature that weights preferentially the low temperature regions from the \ion{He}{1}~lines. This temperature can be determined with the \hecs~program, which uses an analysis involving a maximum likelihood method (see Section~\ref{HeI}), which finds the best simultaneous fit to $T$(\ion{He}{1}), $n_e$, $\tau_{3889}$, and He$^+$/H$^+$. This $T$(\ion{He}{1}) can be combined with the temperatures derived from the oxygen CELs to derive \tc~and $T_0$. In positions 1 and 3 of NGC 456, we simply adopted the \tc~value obtained from the \ion{He}{1}~lines: \tc=0.035$\pm$0.032 and \tc=0.040$\pm$0.040, respectively. For NGC 456-2 and NGC 460-1 the values we obtained with the \hecs~program are \tc=0.053$\pm$0.017 and \tc=0.032$\pm$0.032, respectively.

We used the temperature derived from the ratio of the multiplet 1 of \ion{O}{2} lines to the CELs of [\ion{O}{3}] as given by equations [8]$-$[12] in the work of \citep{pea05},
\begin{equation} T_{{\rm(O II/[O III])}} = T_{4651/5007} = f_1 (T_0,t^2).
\label{TOrec}
\end{equation}

With the oxygen RLs in NGC 456-2 we obtained that $T_0=$10,120 K and \tc=0.083$\pm$0.019 using the equations above. For the case of NGC 460-1, $T_0$=11,370 K and \tc=0.041$\pm$0.027 also with oxygen RLs. In NGC 456 - 2 the oxygen is about 80\% twice ionized (see Section~\ref{IonAbCELs}), hence the value of \tc(He$^+$) $\approx$ \tc(O$^{++}$). NGC 460 has approximately 60\% of its oxygen twice ionized, but since we only have one other \tc~determination to compare it with, we will also assume that \tc(He$^+$) $\approx$ \tc(O$^{++}$).

The combined values of thermal inhomogeneities from the \hecs~program and from the oxygen RLs resulted in \tc=0.067$\pm$0.013 for  NGC 456 Position 2 and \tc=0.036$\pm$0.027 for NGC 460 Position 1.

We were able to obtain the C$^{++}$ ionic abundance from both UV CELs and from optical RLs (see Section \ref{IonaAb}). We found that the abundances are not consistent with a homogeneous temperature structure. A possible value for \tc(C$^{++}$) in NGC 456 is 0.08$\pm$0.04; this value represents the \tc~of the UV observed region.

The \tc~determinations in this paper, with the exception of NGC 456-2, lie within typical values of thermal inhomogeneities measured for Galactic \Hiirs~which range between 0.03 and 0.04 \citep{gar07}. The \tc~determination of NGC 456 Position 2 lies within typical values for extragalactic \Hiirs, which range between 0.03 and 0.11. This is probably because in regions that are farther away it is impossible to isolate ``simple'' volumes and one observes the sum of many physical processes: shadows, high density knots, shock waves, ionization fronts, photoionization, etc.

\section{Ionic Abundances}\label{IonaAb}
\subsection{Helium}\label{HeI}
We used the \hecs~program to obtain the ionic abundance of He$^{+}$, presented in Table~\ref{trionic}. This program is described in \citet{pea11} and it uses a maximum likelihood method to search for the set of parameters ($T_0$, \tc, $n_e$, $\tau_{3889}$, $T$(\ion{He}{1}), and He$^+$/H$^+$) that produce the optimal simultaneous fit to all the observed helium lines as well as the measured $T_e$[\ion{O}{3}] and $T_e$[\ion{O}{2}]. The effective recombination coefficients for the \ion{H}{1} and \ion{He}{1} used were those given by \citet{sto95} for H and by \citet{ben99} and \citet{por07} for He. The collisional contribution was estimated from \citet{saw93} and \citet{kin95}. The optical depth effects in the triplets were estimated with calculations made by \citet{ben02}.

Correction for underlying stellar absorption is also important for helium lines, in a similar way as is for hydrogen lines. Hence, to correct for underlying \ion{He}{1} absorption for lines with $\lambda <$5000 \AA~we used values determined by \citet{gon99}, and for redder lines we used the same values as in \citet{pea05}.

Between 9 and 12 \ion{He}{1} lines were used as input in the \hecs~program to determine the \tc~value for each observed position in both NGC 456 and NGC 460.

The \tc~derived from \ion{He}{1} lines can be used to redetermine abundances in the high-temperature regions considering temperature inhomogeneities, due to the similar ionization potentials of He$^+$ and O$^{++}$. Since the \ion{He}{2} line $\lambda$ 4686 was not detected, the fraction of He$^{++}$/He$^{+}$ is smaller than 1$\times 10^{-3}$ for NGC 456 Position 1, 4$\times 10^{-4}$ for NGC 456 Position 2, and 3$\times 10^{-3}$  for NGC 456 Position 3 and NGC 460 Position 1.

\subsection{C and O from Recombination Lines}\label{CORL}
The ionic abundance for C$^{++}$ presented in Table~\ref{trionic}, was obtained from the measurement of the $\lambda$ 4267 line of \ion{C}{2}. The effective recombination coefficient used was that determined by \citet{dav00} for Case A and with $T=$10,000 K. Although we did measure the $\lambda$ 4267 line in positions 1 and 3 of NGC 456 and in Position 1 of NGC 460, the errors were too large to provide any information other than an upper limit.

The effective recombination coefficients for determining the ionic abundance of oxygen from the lines of the multiplet 1 of \ion{O}{2} were taken from \citet{pei93} and \citet{sto94} assuming Case B for $T_e=$10,000 K and $N_e=$100 cm$^{-3}$. The multiplet consists of eight lines each of which depends on the electron density even though the sum of their intensities, $I$(sum), does not \citep{rui03,bas06,pea10}. Since these lines are very faint, it is often necessary to estimate the unobserved and/or blended lines.

The measurements of the \ion{O}{2} lines in NGC 456-2 and NGC 460-1 are presented in Tables~\ref{tlines} and \ref{tlines2}; the abundances derived from these lines are also presented in Table~\ref{trionic}. The errors in the measurements of the oxygen RLs in positions 1 and 3 of NGC 456 were too large to provide useful information. For the first two mentioned cases, we were able to detect four of the eight lines in the multiplet. Due to the spectral resolution of the observations, those lines were blended into two pairs: $\lambda \lambda$4639$+$42 and $\lambda \lambda$4649$+$51.

Measurements of RLs of heavy elements are not common in long-slit spectrophotometry. Contamination from blends of other lines of other ions could be a concern. To check this, we looked into echelle observations of \Hiirs~with good S/N \citep{pea03,est04,gar04,gar07}; the only other lines that appear in this region of the spectra are: \ion{N}{3} $\lambda \lambda$ 4641 and \ion{N}{2} $\lambda$ 4643. The contribution of the sum of these two N lines compared to the sum of the four O lines observed is about 10\% for objects with metallicity and N/O close to solar, and less than 4\% for 30 Doradus that has N/O about three times less than solar and slightly higher than the SMC; given our S/N, we will ignore it (see Figure \ref{zoomOII}).

\subsection{Ionic Abundances from Collisionally Excited Lines}\label{IonAbCELs}
The ionic carbon abundance from CELs was determined with UV spectra of NGC 456 obtained from the on-line data presented by \citet{bon95}. We measured the [\ion{C}{3}] $\lambda$1907 \ion{C}{3}]$\lambda$1909 lines using the \ion{O}{3}] $\lambda\lambda$1660+66 lines with the UV $C$(H$\beta$) presented in \citet{sea79}, and the atomic data contained in the task \begin{tt}ionic\end{tt}
of IRAF v2.15.1. The value of the C$^{++}$/H$^+$ abundance is presented in Table~\ref{tceionic} with and without considering thermal inhomogeneities.

With the exception of carbon, all other ionic abundances presented in Table~\ref{tceionic} were determined using the task \begin{tt}abund\end{tt}~of IRAF considering only the low and medium-ionization zones, which correspond to the low and high-ionization zones of the present work. For positions 1, 2, and 3 of NGC 456 we used $T_{low}=$13,500 K, 12,400 K, and 12,600 K, respectively; and $T_{high}=$ 12,650 K, 12,165 K, and 12,300 K, respectively.

Ionic abundances considering thermal inhomogeneities, \tc$\neq$0.000, were obtained using the traditional determinations, \tc=0.000, corrected by the formalism presented by \citet{pei69}, see also \citet{pei04}.

\section{Total Abundances}\label{totAbs}
In general we observed that all positions in NGC 456 presented a high O ionization degree implying that there is no substantial amount of He$^0$. The amount of He$^+$ is similar in all positions of both NGC 456 and NGC 460, which may indicate that the ionization degree is also similar in Position 1 of NGC 460.

In the case of C and N, total gaseous abundances were determined with the following equations:
\begin{equation}   \frac{N({\rm C})}{N({\rm H})}={\rm ICF(C)}\frac{N({\rm C^{++}})}{N({\rm H^+})}
\label{totC}
\end{equation}
and
\begin{equation}   \frac{N({\rm N})}{N({\rm H})}={\rm ICF(N)}\frac{N({\rm N^{+}})}{N({\rm H^+})},
\label{totN}
\end{equation}
where we used the ionization correction factor (ICF) for C with respect to O of \citet{gar95}, and multiplied it by the fraction of C$^{++}$/O$^{++}$ in NGC 456-2 following \citet{pea03}; this value amounted to 1.27. The N total gaseous abundance was determined adopting the predicted ICF(N) of \citet{pei69}, $N$(O)/$N$(O$^+$).

The total gaseous abundances for Cl, S, and Ar were determined with the following equations:
\begin{equation}   \frac{N({\rm Cl})}{N({\rm H})}={\rm ICF(Cl)}\frac{N({\rm Cl^{++}})}{N({\rm H^+})},
\label{totCl}
\end{equation}
\begin{equation}   \frac{N({\rm S})}{N({\rm H})}={\rm ICF(S)}\frac{N({\rm S^{+}})+N({\rm S^{++}})}{N({\rm H^+})},
\label{totS}
\end{equation}
and
\begin{equation}   \frac{N({\rm Ar})}{N({\rm H})}={\rm ICF(Ar)}\frac{N({\rm Ar^{++}})+N({\rm Ar^{+++}})}{N({\rm H^+})};
\label{totAr}
\end{equation}
where the corresponding ICFs were estimated from \citet{gar89} for Cl, in the case of S from \citet{pea05}, and for Ar the ICF was taken from \citet{liu00}. For NGC 456-2 these values amounted to 1.26, 1.36, and 1.30, respectively; and 1.26, 1.12, and 1.71 for NGC 460-1.

Following \citet{pei69}, total gaseous abundances of O and Ne were calculated with the equations:
\begin{equation}   \frac{N({\rm O})}{N({\rm H})}=\frac{N({\rm O^{+}})+N({\rm O^{++}})}{N({\rm H^+})}
\label{totO}
\end{equation}
and
\begin{equation}
\frac{N({\rm Ne})}{N({\rm H})}=\frac{N({\rm O^{+}})+N({\rm O^{++}})}{N({\rm O^{++}})}\times \frac{N({\rm Ne^{++}})}{N({\rm H^+})}.
\label{totNe}
\end{equation}

Total gaseous abundances for all available elements are presented in Tables~\ref{tabund4}~and \ref{tabund53} for NGC 456 positions 1, 2, and 3, and in Table~\ref{tabund8} for NGC 460 Position 1. All these abundances are corrected by depletion of O in dust according to \citet{pea10}. Within the errors, abundances in all three positions of NGC 456 are very similar. Observations with higher S/N of NGC 456-1, -3 and of NGC 460-1 would allow to measure a more accurate value of \tc.

According to \citet{pea10}, a correction of 0.10 dex should be made to the total O abundances due to depletion of oxygen in dust. The adopted total abundance values for all available elements are presented in Tables~\ref{tta} and \ref{ttta}. For comparison, we also present protosolar abundances and abundances of the Orion nebula, NGC 6822, and NGC 346, which is the brightest \Hiir~of the SMC \citep{pei00}. In addition to this, we obtained an abundance determination for the ISM of the solar vicinity from the Galactic gradient derived from \Hiirs, that amounted to 12+log(O/H)$\,$=$\,$8.81 once corrected by 0.12 for dust depletion \citep{est05}.

\section{A Systematic Correction to the Direct Method}\label{clasMR}
\subsection{The Direct Method and its Shortcomings}\label{DMshort}
The ADF is defined as the ratio of abundances determined using RLs vs. abundances obtained with CELs. ADFs are now well established for both Planetary Nebulae and \Hiirs. RLs have a temperature dependence given approximately by $I\varpropto T_e^{-1}$, thus are brighter at lower temperatures; whereas CELs have a much stronger dependence, $I\varpropto exp($k$ T_e/\Delta E)T_e^{-1/2}$, and thus brighter at higher temperatures.

Several studies \citep[][and references therein]{pei93,pei95,tsa03,tsa04,pea05,gar07,pei07,est09} have shown that ADF typical values are from 1.5 to 3 for most \Hiirs, and from 1.5 to 5 for most Planetary Nebulae. This is clear evidence that the traditional or direct method of abundance determinations -- involving the assumption of homogeneous temperature determined from the emission line ratio 4363/(4959+5007) \AA~of [\ion{O}{3}] -- needs to be corrected. Since the work of \citet{pei67} and \citet{pei69}, temperature inhomogeneities have been presented as a viable solution to the so called ADF problem. \citet{pei11} discuss several methods that have been used to obtain \tc~values based on: i) the comparison between the temperatures derived from the ratio of the Balmer and Paschen continua to the Balmer line intensities and the temperatures derived from CELs; ii) the comparison of the C$^{++}$ abundances derived using [\ion{C}{3}]+\ion{C}{3}] with abundances obtained from \ion{C}{2}; iii) the comparison of O$^{++}$ abundances derived with [\ion{O}{3}] with abundances obtained using \ion{O}{2}; iv) the comparison of the temperature that can be derived from \ion{He}{1} lines to the temperatures derived from CELs; v) a high spatial resolution map of 1.5$\times10^6$ columnar temperatures in the Orion nebula determined by \citet{ode03}; and vi) the comparison of the different ways to calibrate Pagel's method. In all cases, the \tc~values are consistent with each other and the ADF problem goes away.

In the case of photoionization models, the intensity of nebular lines (such as $\lambda$ 3727 [\ion{O}{2}], or $\lambda$ 6584 [\ion{N}{2}]) can be adjusted in most models of both Planetary Nebulae and \Hiirs; however, auroral lines (such as $\lambda$ 4363 [\ion{O}{3}]) are in general not adjusted in those photoionization models.

\subsection{Typical Correction for Low Metallicity \Hiirs}\label{DMtypcor}
We find that O/H abundances in \Hiirs~determined with the direct method are underestimated and have to be corrected by factors of 0.25$\,$-$\,$0.45 dex. The correction we propose is due to two distinct and important physical processes: thermal inhomogeneities and depletion of oxygen into dust grains.

When allowing thermal inhomogeneities to exist in the abundance determination process, we find that the derived oxygen abundances in \Hiirs~increase by 0.15$\,$-$\,$0.35 dex, depending on the particular characteristics of the thermal structure of each photoionized region.

According to \citet{pea10}, abundances of photoionized regions require a correction of 0.09$\,$-$\,$0.10 dex for low metallicity \Hiirs~due to depletion of oxygen into dust. Therefore the total correction amounts to about 0.35 dex in the total O abundance.

Many objects have \tc~determined with uncertainties larger than about 0.030, and frequently these determinations are consistent with 
\tc=0.000; the fact that many such objects exist does not mean that the \tc=0.000 or that there are many objects with negligible 
temperature inhomogeneities. Since all the objects where the determinations have uncertainties smaller than 0.015, are not
consistent with a homogeneous temperature, it means that most objects have meaningful temperature inhomogeneities. To ignore the 
presence of these inhomogeneities because the data quality is not good enough, produces a very large bias in the study of 
\Hiirs~in general. For this reasons we recommend to use this average correction of 0.35 dex in the total O abundance, and to explicitly present it so that, if the recommended correction improves with time, it is clear what to do.

\subsection{Implications for the Determinations of Extragalactic Chemical Abundances}\label{DMimp}
Studies of abundances in \Hiirs~can also be used to check our understanding of the chemical evolution of our own Galaxy. \citet{pei11} and \citet{car11} find that careful determination of chemical abundances in \Hiirs~(once corrected for depletion of oxygen into dust and for temperature inhomogeneities), are consistent with: i) the protosolar abundances after correcting them for the Galactic chemical evolution \citep{asp09,car11}, ii) young F and G stars of the solar vicinity \citep{bens06}, and iii) O and B stars of the Orion region \citep{pry08,sim11}.

Since the afore mentioned methods of determining abundances for stars and \Hiirs~are entirely independent, the consistency in the results allows to individually validate each method. In particular it implies that a model with temperature inhomogeneities is a better approximation to reality than the direct method, that considers a homogeneous temperature structure.

Almost all studies of \Hiirs~in the literature are made assuming homogeneous, or nearly homogeneous, temperature throughout the whole object, hence, the abundances derived from these works are underestimated by a factor of approximately 2. The study of abundances in low metallicity \Hiirs~is important to set limits in the models of various areas of astrophysics such as the radial abundance gradients in spiral galaxies \citep{vila92,zar94,est05}, the chemical evolution of starburst galaxies \citep[e.g.][]{liu08,per08,bor11,car11,pei11}, the mass-luminosity relation and the mass-metallicity relation \citep[e.g.][]{tre04,kew08,man10,thu10}, the initial mass function, star formation rate and yield determinations \citep[e.g.][]{car06,pil07,pei11}, and the primordial helium abundance \citep[][and references therein]{pei07,izo10,pei10}, among others. Consequently, it is of great importance for these studies not to use abundances derived from the direct method, and to use abundances obtained with a method that either allows the presence of thermal inhomogeneities or is not affected by them (e.g. CELs corrected by temperature inhomogeneities or RLs).

\section{Discussion: the Thermal Structure of Gaseous Nebulae}\label{disc}
In photoionization models, the temperature is nearly homogeneous because the processes considered for both heating and cooling are nearly proportional to the density squared. Strong thermal inhomogeneities require processes that generate either heating or cooling that are not proportional to the density squared.

Observationally, the ADF problem emerges when a homogeneous temperature is assumed. \citet{ode03} presented a high spatial resolution map of the columnar electron temperatures of the Trapezium with which they found a \tc~value from different lines of sight within the nebula, whereas \citet{tsa11a}, studying components and different velocities, found that each line of sight has thermal inhomogeneities.

All available explanations in the literature for the presence of the ADF require the presence of thermal inhomogeneities. There are two families of mechanisms that produce temperature inhomogeneities: i) an inhomogeneous cooling function of the photoionized region, i.e. chemical inhomogeneities \citep{tor90,ten96,liu00,tsa05,erc07,sta07}, and ii) an inhomogeneous heating, e.g. shock waves, advancing photoionization fronts, shadowed regions, magnetic reconnection \citep[][and references therein]{pei06}. There is a small fraction of Planetary Nebulae and \Hiirs~that are chemically inhomogeneous; these objects will not be discussed in this paper. Models that are chemically inhomogeneous predict abundances that are intermediate between those derived with the direct method and from RLs. For most objects an inhomogeneous heating mechanism is in better agreement with abundances derived from stellar objects (see section \ref{DMimp}), and with the thermal structure of the Orion nebula (see section \ref{DMshort}). Moreover, \tc~values derived with independent methods are consistent with each other within the errors. Specifically, the fact that the \tc~values from He and O are consistent, implies that the material in the nebulae is well mixed since these elements have different formation histories \citep{car11}.

Typical photoionization model values of \tc~are about one order of magnitude smaller than the temperature inhomogeneities determined from observations. With near photoionized objects such as the Orion nebula, extremely detailed studies can be made to measure precise values of \tc: e.g. \citet{ode03} presented a high spatial resolution map of the columnar electron temperatures of the Trapezium with which they found that shadowed regions can significantly contribute to the observed thermal inhomogeneities; \citet{tsa11a} and \citet{tsa11b} were able to isolate individual high density regions and found that CELs underestimate abundances; and \citet{mes09a} and \citet{mes09b} studied the head of HH 202 and found that the ionization front has a \tc~value significantly higher than that of the nebula.

Photoionization models include an energy injection proportional to the density squared; this should be a good approximation if the energy were dominated by photoionization. However, the measurements of the filling factor (which is defined as the occupied volume over the total volume) argue against such an energy budget model. Typical filling factor values for \Hiirs~are in the 0.01-0.10 range. These values are showing that \Hiirs~are pretty much empty, but they are not showing the origin of the energy that is keeping the material apart. This is strong evidence that the traditional energy budget model to study \Hiirs~is incomplete. The fact that it is not known what is feeding the filling factor in \Hiirs~does not mean that abundance results considering \tc$>$0.00 are incorrect, it means that models have to be improved to better understand the physics and astrophysics of these objects. Temperature inhomogeneities are showing that photoionization models have to be corrected to account for the additional physics that has not been included.

Chemical abundances determined using CELs strongly depend on the thermal structure of the nebula, whereas abundances obtained with RLs do not. Even though we know many processes capable of generating temperature inhomogeneities, we do not know, quantitatively, the relative importance of each one of them. We can estimate the magnitude and the effect that these thermal inhomogeneities have on the chemical abundances. Although it is interesting to study the origin of such thermal inhomogeneities, it is beyond the scope of this paper to do so. In order to determine abundances we need to know the value of \tc~and not what is producing it.

\section{Conclusions}\label{conc}
A detailed analysis was performed for the SMC \Hiirs~NGC 456 and NGC 460. This analysis involves abundance determinations considering thermal inhomogeneities  (\tc$\neq$0.00); for the purpose of comparison with other authors, we also determined abundances with the direct or traditional method, which assumes a homogeneous temperature structure (\tc=0.00). The abundances of NGC 346 \citep{pei00}, NGC 456, and NGC 460, are similar once corrected for dust depletion. The O/H abundances of the SMC \Hiirs~are 0.58 dex lower than those of the ISM of the solar vicinity, and 0.46 dex lower than the solar photosphere. This indicates that the Small Magellanic Cloud is chemically less evolved than the Milky Way and so is the best laboratory to study in detail the ISM of low metallicity galaxies.

Determinations of oxygen abundances in this work were done with three methods: i) using RLs, ii) using the direct method for CELs, and iii) using CELs relaxing the homogeneous temperature assumption. The first two methods are not compatible. Temperature inhomogeneities of some sort are required to reconcile observations of RLs and CELs. The ADF only becomes a problem if one demands a homogeneous thermal structure from the start. Our results confirm that the ADF problem can be solved by taking into account the presence of thermal inhomogeneities in a chemically homogeneous medium.

Abundances of \Hiirs~are used to constrain models of many areas of astrophysics, therefore, we need to derive abundances from gaseous nebulae with the highest precision possible. A crucial aspect of the available observations is that they require a complex thermal structure; consequently, thermal inhomogeneities must be taken into account to determine abundances.

There are many ways to obtain a \tc~value \citep{pei11}, the most commonly used is based on the comparison of abundances derived with [\ion{O}{3}] lines with abundances derived using \ion{O}{2} lines. In this paper we also obtained \tc~values from \ion{He}{1} lines and from the comparison of abundances derived using IUE UV \ion{C}{3}] and [\ion{C}{3}] lines with abundances derived from optical \ion{C}{2} lines. For both NGC 456 and NGC 460, we combined \tc~values of O$^{++}$ and He$^{+}$ to obtain an average \tc. The \tc~value for C$^{++}$ in NGC 456 was not included in such average of \tc~because optical and UV observations do not come from the same place in the \Hiir; nonetheless, this \tc(C$^{++}$) is relevant because it is not consistent with a homogeneous temperature structure in the \Hiir.

To obtain accurate values of \tc, good S/N is required for the faint lines of \ion{He}{1}, \ion{C}{2}, and \ion{O}{2}. \ion{He}{1} and \ion{O}{2} for the same volume are consistent within the errors. Since each \tc~value comes from a different element and they have different formation histories, this is evidence that the material within the nebula is well mixed; otherwise those \tc~values of different elements would differ from one another.

The homogeneous temperature assumption of the direct method should only be taken as a first approximation because, in addition to photoionization, other energy sources have to be taken into account. Temperature inhomogeneities measure the overall importance of the many mechanisms that inject energy into the system. We find that O/H abundances of low metallicity \Hiirs~are systematically underestimated and have to be corrected by a factor of 0.25$\,$-$\,$0.45 dex. We suggest that, if no additional information is available, an increase of 0.35 dex on the O/H ratio should be used; this correction includes temperature inhomogeneities and oxygen depletion into dust.

We are grateful to the anonymous referee for the careful reading of the manuscript and several excellent suggestions. MAPG and AP received partial support from UNAM (grant PAPIIT 112911), and AP and MP received partial support from CONACyT (grant 129753). MTR received partial support from Proyecto Basal PB06 (CATA) and Centro de Astrofísica FONDAP.

\clearpage

\begin{deluxetable}{lcr@{--}lcc} 
\tablecaption{Journal of Observations
\label{tobs}}
\tablewidth{0pt}
\tablehead{
\colhead{Grism}  & 
\colhead{Filter} &
\multicolumn{2}{c}{$\lambda$ (\AA)} &
\colhead{Resolution ($\lambda$/$\Delta \lambda$)} &
\colhead{Exp. Time (s)}}
\startdata
GRIS-600B+12 &   -   & 3450 &5900  & 1300 & 3$\times$720  \\
GRIS-600R+14 & GG435 & 5250 &7450  & 1700 & 3$\times$600  \\
GRIS-300V    & GG375 & 3850 &8800  &  700 & 3$\times$120  \\
\enddata
\end{deluxetable} 
\clearpage

\begin{deluxetable}{ccccc}
\tablecaption{Absolute Equivalent Widths
\label{tEWs}}
\tablewidth{0pt}
\tablehead{
\multicolumn{2}{c}{Hydrogen} && \multicolumn{2}{c}{Helium} \\
\cline{1-2}
\cline{4-5}
\colhead{Line} & \colhead{$\frac{\rm EW_{abs}(\rm Line) }{\rm EW_{abs}(\rm H\beta)}$}
&& \colhead{Line} & \colhead{$\frac{\rm EW_{abs}(\rm Line) }{\rm EW_{abs}(\rm H\beta)}$}
}
\startdata
H$\alpha$\tablenotemark{a}	&0.90	&& 3820\tablenotemark{a}	&0.108\\
H$\gamma$\tablenotemark{a}	&1.05	&& 4026\tablenotemark{a}	&0.100\\
H$\delta$\tablenotemark{a}	&1.08	&& 4388\tablenotemark{a}	&0.084\\
H7\tablenotemark{a}		&0.99	&& 4471\tablenotemark{a}	&0.179\\
H8\tablenotemark{a,b}		&0.93	&& 4922\tablenotemark{a}	&0.107\\
H9\tablenotemark{a}		&0.78	&& 4009\tablenotemark{d}	&0.100\\
H10\tablenotemark{c}		&0.67	&& 5016\tablenotemark{d}	&0.114\\
H11\tablenotemark{c}		&0.54	&& 5048\tablenotemark{d}	&0.115\\
H12\tablenotemark{c}		&0.42	&& 5876\tablenotemark{d}	&0.138\\
H13\tablenotemark{c}		&0.35	&& 6678\tablenotemark{d}	&0.082\\
H14\tablenotemark{c}		&0.30	&& 7281\tablenotemark{d}	&0.032\\
H15\tablenotemark{c}		&0.25	&&\\
H16\tablenotemark{c}		&0.20	&&\\
H17\tablenotemark{c}		&0.16	&&\\
H18\tablenotemark{c}		&0.13	&&\\
H19\tablenotemark{c}		&0.10	&&\\
H20\tablenotemark{c}		&0.08	&&\\
H21\tablenotemark{c}		&0.06	&&\\
H22\tablenotemark{c}		&0.05	&&\\
\enddata
\tablenotetext{a} {\citet{gon99}.}
\tablenotetext{b} {Note that \ion{He}{1}(3889) is included with H8.}
\tablenotetext{c} {Extrapolations based on VLT echelle observations of 30 Doradus \citep{pea03}.}
\tablenotetext{d} {M. Cervi\~no, private conversation.}
\end{deluxetable}
\clearpage

\begin{deluxetable}{lcr@{}lr@{}lr@{}lccr@{}lr@{}lccr@{}lr@{}lc}
\tabletypesize{\scriptsize}
\tablewidth{0pt}
\tablecaption{NCG 456: Line Intensities for Positions 1, 2, and 3
\label{tlines}}
\tablehead{
&&&& \multicolumn{5}{c}{Position 1} &&\multicolumn{5}{c}{Position 2} &&\multicolumn{5}{c}{Position 3} \\
\cline{5-9}\cline{11-15}\cline{17-21}\\
\colhead{$\lambda$ (\AA)} & \colhead{Id.} & \multicolumn{2}{c}{$f(\lambda$)} &
\multicolumn{2}{c}{$F(\lambda)$} &
\multicolumn{2}{c}{$I(\lambda)$} &
\colhead{\%~err.} &&
\multicolumn{2}{c}{$F(\lambda)$}&
\multicolumn{2}{c}{$I(\lambda)$} &
\colhead{\%~err.} &&
\multicolumn{2}{c}{$F(\lambda)$} &
\multicolumn{2}{c}{$I(\lambda)$} &
\colhead{\%~err.} }
\startdata

3587    &\ion{He}{1}   &0.&272&    0.&56&   0.&64& 18&&   0.&59&   0.&66& 10&&   0.&93& 1.&02&   28\\
3614    &\ion{He}{1}   &0.&271&    0.&95&   1.&08& 14&&   1.&09&   1.&22&  7&&   0.&91& 0.&99&   29\\
3634    &\ion{He}{1}   &0.&270&    0.&89&   1.&01& 15&&   1.&23&   1.&37&  7&&  \mcnd &  \mcnd &\nd\\
3676    &H\,22         &0.&269&   \mcnd &  \mcnd &\nd&&   0.&35&   0.&40& 13&&  \mcnd &  \mcnd &\nd\\
3679    &H\,21         &0.&268&   \mcnd &  \mcnd &\nd&&   0.&43&   0.&48& 11&&  \mcnd &  \mcnd &\nd\\
3683    &H\,20         &0.&267&   \mcnd &  \mcnd &\nd&&   0.&38&   0.&43& 12&&   0.&68&   0.&76& 33\\
3687    &H\,19         &0.&266&   \mcnd &  \mcnd &\nd&&   0.&54&   0.&61& 10&&   0.&64&   0.&72& 34\\
3692    &H\,18         &0.&265&   \mcnd &  \mcnd &\nd&&   0.&63&   0.&71&  9&&   0.&55&   0.&62& 37\\
3695    &H\,17         &0.&264&   \mcnd &  \mcnd &\nd&&   0.&76&   0.&99&  9&&   0.&99&   1.&35& 28\\
3704    &H\,16         &0.&262&   \mcnd &  \mcnd &\nd&&   1.&24&   1.&56&  7&&   1.&18&   1.&62& 25\\
3712    &H\,15         &0.&260&   \mcnd &  \mcnd &\nd&&   1.&22&   1.&58&  7&&   1.&55&   2.&11& 22\\
3722    &H\,14+[\ion{S}{3}]
		       &0.&257&    1.&54&   1.&99& 11&&   1.&60&   2.&05&  6&&   1.&46&   2.&11& 23\\
3726    &[\ion{O}{2}]  &0.&256&   65.&02&  73.&16&  2&&  58.&79&  65.&23&  1&&  61.&13&  66.&59&  4\\
3729    &[\ion{O}{2}]  &0.&255&   86.&89&  97.&72&  2&&  73.&92&  81.&98&  1&&  83.&15&  90.&54&  3\\
3734    &H\,13         &0.&254&    2.&13&   2.&69&  9&&   1.&70&   2.&20&  6&&   1.&69&   2.&44& 21\\
3750    &H\,12         &0.&250&    2.&38&   3.&00&  9&&   2.&48&   3.&11&  5&&   2.&09&   2.&97& 19\\
3770    &H\,11         &0.&245&    3.&14&   3.&97&  8&&   3.&24&   4.&06&  4&&   3.&05&   4.&20& 16\\
3798    &H\,10         &0.&238&    4.&11&   5.&14&  7&&   4.&31&   5.&34&  4&&   3.&95&   5.&39& 14\\
3820    &\ion{He}{1}   &0.&233&    0.&75&   0.&92& 16&&   0.&80&   0.&97&  8&&   1.&65&   1.&91& 21\\
3836    &H\,9          &0.&229&    5.&97&   7.&29&  6&&   5.&98&   7.&26&  3&&   5.&48&   7.&19& 12\\
3867    &[\ion{Ne}{3}] &0.&222&   17.&91&  19.&82&  3&&  22.&23&  24.&30&  2&&  22.&82&  24.&55&  6\\
3889    &H\,8+\ion{He}{1}
                       &0.&218&   16.&71&  19.&25&  3&&  16.&57&  18.&93&  2&&  17.&26&  20.&04&  7\\
3967    &[\ion{Ne}{3}] &0.&201&   20.&25&  22.&68&  3&&  22.&15&  24.&01&  2&&  21.&25&  22.&68&  6\\
4009    &\ion{He}{1}   &0.&193&   \mcnd &  \mcnd &\nd&&   0.&17&   0.&69& 18&&  \mcnd &  \mcnd &\nd\\
4026    &\ion{He}{1}   &0.&190&    1.&37&   1.&58& 12&&   1.&60&   1.&90&  6&&   1.&71&   1.&97& 21\\
4067    &[\ion{S}{2}]  &0.&182&    1.&01&   1.&10& 14&&   0.&93&   0.&99&  8&&   0.&71&   0.&76& 33\\
4076    &[\ion{S}{2}]  &0.&181&    0.&32&   0.&35& 24&&   0.&30&   0.&32& 14&&  \mcnd &  \mcnd &\nd\\
4102    &H$\delta$     &0.&176&   23.&81&  26.&65&  3&&  23.&82&  26.&45&  2&&  23.&62&  26.&53&  6\\
4125    &[\ion{Fe}{2}] &0.&172&   \mcnd &  \mcnd &\nd&&   0.&30&   0.&32& 14&&  \mcnd &  \mcnd &\nd\\
4144    &\ion{He}{1}   &0.&169&   \mcnd &  \mcnd &\nd&&   0.&30&   0.&42& 14&&  \mcnd &  \mcnd &\nd\\
4169    &\ion{O}{2}    &0.&164&   \mcnd &  \mcnd &\nd&&   0.&18&   0.&19& 18&&  \mcnd &  \mcnd &\nd\\
4185    &\ion{O}{2}    &0.&161&   \mcnd &  \mcnd &\nd&&   0.&12&   0.&13& 21&&  \mcnd &  \mcnd &\nd\\
4267    &\ion{C}{2}    &0.&143& $<$0.&30&$<$0.&30&\nd&&   0.&04&   0.&04& 40&& $<$0.&30& $<$0.&30&\nd\\
4340    &H$\gamma$     &0.&128&   45.&03&  48.&25&  2&&  44.&52&  47.&51&  1&&  44.&29&  47.&28&  4\\
4363    &[\ion{O}{3}]  &0.&122&    4.&12&   4.&34&  7&&   4.&36&   4.&57&  4&&   4.&22&   4.&38& 13\\
4388    &\ion{He}{1}   &0.&116&    0.&58&   0.&66& 18&&   0.&42&   0.&50& 12&&  \mcnd &  \mcnd &\nd\\
4471    &\ion{He}{1}   &0.&094&    3.&77&   4.&13&  7&&   3.&72&   3.&97&  4&&   3.&89&   4.&20& 14\\

4563    &\ion{Mg}{1}]  &0.&070&   \mcnd &  \mcnd &\nd&&   0.&08&   0.&08& 27&&  \mcnd &  \mcnd &\nd\\
4571    &\ion{Mg}{1}]  &0.&068&   \mcnd &  \mcnd &\nd&&   0.&05&   0.&05& 33&&  \mcnd &  \mcnd &\nd\\
4639+42 &\ion{O}{2}    &0.&051&   \mcnd &  \mcnd &\nd&&   0.&06&   0.&06& 30&&  \mcnd &  \mcnd &\nd\\
4649+51 &\ion{O}{2}    &0.&049&   \mcnd &  \mcnd &\nd&&   0.&07&   0.&07& 28&&  \mcnd &  \mcnd &\nd\\
4658    &[\ion{Fe}{3}] &0.&047&    0.&36&   0.&36& 23&&   0.&29&   0.&30& 14&&  \mcnd &  \mcnd &\nd\\
4701    &[\ion{Fe}{3}] &0.&037&    0.&10&   0.&11& 43&&   0.&10&   0.&10& 24&&  \mcnd &  \mcnd &\nd\\
4711    &[\ion{Ar}{4}]+\ion{He}{1}
                       &0.&034&    0.&44&   0.&53& 21&&   0.&50&   0.&59& 11&&  \mcnd &  \mcnd &\nd\\
4740    &[\ion{Ar}{4}] &0.&028&   \mcnd &  \mcnd &\nd&&   0.&06&   0.&06& 31&&  \mcnd &  \mcnd &\nd\\
4755    &[\ion{Fe}{3}] &0.&024&   \mcnd &  \mcnd &\nd&&   0.&07&   0.&07& 29&&  \mcnd &  \mcnd &\nd\\
4861    &H$\beta$      &0.&000&  100.&00& 100.&00&  2&& 100.&00& 100.&00&  1&& 100.&00& 100.&00&  3\\
4881    &[\ion{Fe}{3}] &-0.&004&   0.&24&   0.&24& 28&&   0.&10&   0.&10& 23&&  \mcnd &  \mcnd &\nd\\
4922    &\ion{He}{1}   &-0.&013&   1.&10&   1.&14& 13&&   1.&03&   1.&07&  7&&   1.&29&   1.&38& 24\\
4959    &[\ion{O}{3}]  &-0.&021& 116.&26& 114.&42&  2&& 135.&29& 133.&39&  1&& 127.&17& 124.&85&  2\\
4986    &[\ion{Fe}{3}] &-0.&027&   0.&60&   0.&59& 18&&   0.&30&   0.&30& 14&&   \mcnd&  \mcnd &\nd\\
5007    &[\ion{O}{3}]  &-0.&032& 345.&85& 338.&55&  1&& 404.&17& 396.&63&  1&& 375.&64& 367.&27&  2\\
5016    &\ion{He}{1}   &-0.&034&   2.&48&   2.&49&  9&&   2.&44&   2.&45&  5&&   2.&74&   2.&78& 17\\
5041    &\ion{S}{3}    &-0.&040&  \mcnd &  \mcnd &\nd&&   0.&11&   0.&11& 22&&   0.&72&   0.&70& 33\\
5048    &\ion{He}{1}   &-0.&041&   0.&27&   0.&32& 26&&   0.&20&   0.&25& 17&&  \mcnd &  \mcnd &\nd\\
5056    &\ion{S}{3}    &-0.&043&  \mcnd &  \mcnd &\nd&&   0.&08&   0.&07& 27&&  \mcnd &  \mcnd &\nd\\
5192    &[\ion{Ar}{3}] &-0.&072&  \mcnd &  \mcnd &\nd&&   0.&08&   0.&08& 26&&  \mcnd &  \mcnd &\nd\\
5198    &[\ion{N}{1}]  &-0.&074&   0.&21&   0.&20& 30&&   0.&12&   0.&12& 21&&  \mcnd &  \mcnd &\nd\\
5270    &[\ion{Fe}{3}] &-0.&089&   0.&17&   0.&17& 33&&   0.&16&   0.&16& 19&&  \mcnd &  \mcnd &\nd\\
5517    &[\ion{Cl}{3}] &-0.&140&   0.&41&   0.&38& 22&&   0.&43&   0.&41& 11&&   0.&46&   0.&43& 41\\
5537    &[\ion{Cl}{3}] &-0.&144&   0.&33&   0.&30& 24&&   0.&31&   0.&29& 13&&   0.&35&   0.&33& 47\\
5755    &[\ion{N}{2}]  &-0.&191&   0.&22&   0.&20& 30&&   0.&15&   0.&14& 19&&  \mcnd &  \mcnd &\nd\\
5876    &\ion{He}{1}   &-0.&216&  11.&76&  10.&58&  4&&  12.&25&  11.&17&  2&&  11.&74&  10.&82&  8\\
6300    &[\ion{O}{1}]  &-0.&285&  \mcnd &  \mcnd &\nd&&   1.&11&   0.&98&  7&&   1.&99&   1.&77& 20\\
6312    &[\ion{S}{3}]  &-0.&286&   1.&12&   1.&00& 19&&   2.&02&   1.&78&  5&&   1.&12&   1.&00& 26\\
6364    &[\ion{O}{1}]  &-0.&294&  \mcnd &  \mcnd &\nd&&   0.&38&   0.&33& 12&&   2.&96&   2.&62& 16\\
6371    &[\ion{S}{3}]  &-0.&295&  \mcnd &  \mcnd &\nd&&   0.&05&   0.&05& 33&&  \mcnd &  \mcnd &\nd\\
6548    &[\ion{N}{2}]  &-0.&320&   5.&42&   4.&61&  6&&   2.&94&   2.&55&  4&&   2.&68&   2.&35& 17\\
6563    &H$\alpha$     &-0.&322& 339.&15& 288.&06&  1&& 330.&81& 287.&11&  1&& 327.&01& 287.&21&  2\\
6583    &[\ion{N}{2}]  &-0.&324&   12.&06& 10.&83&  4&&   8.&19&   7.&10&  3&&   8.&04&   7.&05& 10\\
6678    &\ion{He}{1}   &-0.&337&   3.&45&   2.&90&  7&&   3.&73&   3.&23&  4&&   3.&81&   3.&37& 14\\
6716    &[\ion{S}{2}]  &-0.&342&  12.&85&  10.&83&  4&&  10.&04&   8.&63&  2&&  10.&34&   9.&00&  9\\
6731    &[\ion{S}{2}]  &-0.&343&   9.&46&   7.&97&  5&&   8.&35&   7.&18&  3&&   7.&20&   6.&26& 10\\
7065    &\ion{He}{1}   &-0.&383&   2.&54&   2.&11&  9&&   2.&95&   2.&51&  4&&   2.&32&   2.&02& 18\\
7136    &[\ion{Ar}{3}] &-0.&391&   8.&87&   7.&30&  5&&   9.&90&   8.&34&  2&&   8.&86&   7.&57&  9\\

7281    &\ion{He}{1}   &-0.&406&   0.&71&   0.&59& 16&&   0.&75&   0.&63&  9&&  \mcnd &  \mcnd &\nd\\
7319    &[\ion{O}{2}]  &-0.&410&   3.&65&   2.&97&  7&&   3.&79&   3.&16&  4&&   2.&87&   2.&43& 16\\
7330    &[\ion{O}{2}]  &-0.&411&   2.&98&   2.&43&  8&&   3.&03&   2.&53&  4&&   2.&45&   2.&08& 18\\
7442    &\ion{N}{2}    &-0.&422&   0.&24&   0.&20& 28&&   0.&25&   0.&21& 15&&   \mcnd & \mcnd &\nd\\
7510    &\ion{C}{2}    &-0.&429&   0.&82&   0.&67& 15&&   1.&06&   0.&88&  7&&   1.&03&   0.&87& 27\\
7552    &[\ion{Fe}{2}] &-0.&440&   0.&81&   0.&65& 15&&   0.&88&   0.&72&  8&&   0.&93&   0.&78& 29\\
7720    &[\ion{Fe}{2}] &-0.&450&   0.&86&   0.&68& 15&&   1.&14&   0.&94&  7&&  \mcnd &  \mcnd &\nd\\
7751    &[\ion{Ar}{3}] &-0.&452&   2.&02&   1.&62& 10&&   2.&05&   1.&68&  5&&   0.&99&   0.&83& 28\\
7994    &\ion{N}{2}    &-0.&473&   1.&35&   1.&07& 12&&   1.&38&   1.&12&  6&&  \mcnd &  \mcnd &\nd\\
8360    &P\,22         &-0.&503&  \mcnd &  \mcnd &\nd&&   0.&14&   0.&12& 20&&  \mcnd &  \mcnd &\nd\\
8375    &P\,21         &-0.&504&  \mcnd &  \mcnd &\nd&&   0.&11&   0.&08& 23&&  \mcnd &  \mcnd &\nd\\
8413    &P\,19         &-0.&507&  \mcnd &  \mcnd &\nd&&   0.&22&   0.&18& 16&&  \mcnd &  \mcnd &\nd\\
8467    &P\,17         &-0.&512&  \mcnd &  \mcnd &\nd&&   0.&36&   0.&29& 12&&  \mcnd &  \mcnd &\nd\\
8487    &[\ion{Fe}{2}] &-0.&514&  \mcnd &  \mcnd &\nd&&   1.&91&   1.&53&  5&&   2.&60&   2.&12& 17\\
8502    &P\,16+[\ion{C}{3}]
                       &-0.&515&  \mcnd &  \mcnd &\nd&&   0.&51&   0.&41& 10&&  \mcnd &  \mcnd &\nd\\
8546    &P\,15+[\ion{Cl}{3}]
                       &-0.&518&   0.&43&   0.&33& 21&&   0.&45&   0.&36& 11&&   1.&57&   1.&28& 22\\
8599    &P\,14         &-0.&523&   0.&34&   0.&26& 24&&   0.&58&   0.&46& 10&&  \mcnd &  \mcnd &\nd\\
8665    &P\,13         &-0.&528&   2.&59&   1.&99&  9&&   0.&80&   0.&64&  8&&   1.&43&   1.&16& 23\\
8694    &\ion{N}{2}+[\ion{Ne}{2}]
                       &-0.&530&  \mcnd	&  \mcnd &\nd&&   0.&13&   0.&10& 21&&  \mcnd &  \mcnd &\nd\\
8751    &P\,12         &-0.&535&  \mcnd &  \mcnd &\nd&&   1.&01&   0.&80&  7&&  \mcnd &  \mcnd &\nd\\

\hline

\multicolumn{2}{l}{$EW$(H$\beta$), in \AA}
&&& \multicolumn{5}{c}{341}&&\multicolumn{5}{c}{383}&&\multicolumn{5}{c}{227}\\

\multicolumn{2}{l}{$EW_{abs}$(H$\beta$), in \AA}
&&& \multicolumn{5}{c}{2.4$\pm$0.1}&&\multicolumn{5}{c}{2.0$\pm$0.1}&&\multicolumn{5}{c}{2.0$\pm$0.1}\\

\multicolumn{2}{l}{$C$(H$\beta$), in dex}
&&& \multicolumn{5}{c}{0.16$\pm$0.03} && \multicolumn{5}{c}{0.19$\pm$0.03} && \multicolumn{5}{c}{0.21$\pm$0.03}\\

\multicolumn{3}{l}{$F$(H$\beta$), in erg s$^{-1}$ cm$^{-2}$}
&&\multicolumn{5}{c}{1.66$\times10^{-13}$} &&\multicolumn{5}{c}{1.04$\times10^{-13}$} &&\multicolumn{5}{c}{2.46$\times10^{-14}$}\\

\multicolumn{3}{l}{$I(H\beta$), in erg s$^{-1}$ cm$^{-2}$}
&&\multicolumn{5}{c}{2.71$\times10^{-13}$} &&\multicolumn{5}{c}{1.60$\times10^{-13}$} &&\multicolumn{5}{c}{3.62$\times10^{-14}$}\\

\enddata

\end{deluxetable}
\clearpage

\begin{deluxetable}{lcr@{}lr@{}lr@{}lcc}

\tabletypesize{\scriptsize}
\tablewidth{0pt}
\tablecaption{NCG 460: Line Intensities of Position 1\label{tlines2}}
\tablehead{
\colhead{$\lambda$ (\AA)} & \colhead{Id.} & \multicolumn{2}{c}{$f(\lambda$)} &
\multicolumn{2}{c}{$F(\lambda)$} &
\multicolumn{2}{c}{$I(\lambda)$} &
\colhead{\%~err.}
}
\startdata

3695    &H\,17         &0.&264&    0.&84&   1.&06& 20\\
3704    &H\,16         &0.&262&    2.&03&   2.&42& 13\\
3712    &H\,15         &0.&260&    1.&68&   2.&07& 14\\
3722    &H\,14+[\ion{S}{3}]
	               &0.&257&    2.&35&   3.&05& 12\\
3726    &[\ion{O}{2}]  &0.&246&   99.&14& 111.&49&  2\\
3729    &[\ion{O}{2}]  &0.&255&  118.&27& 132.&94&  2\\
3734    &H\,13         &0.&254&    1.&82&   2.&52& 14\\
3750    &H\,12         &0.&250&    2.&30&   2.&92& 12\\
3770    &H\,11         &0.&245&    4.&74&   5.&65&  9\\
3798    &H\,10         &0.&238&    5.&25&   6.&28&  8\\
3820    &\ion{He}{1}   &0.&233&    2.&66&   3.&04& 11\\
3836    &H\,9          &0.&229&    6.&29&   7.&49&  7\\
3867    &[\ion{Ne}{3}] &0.&222&   19.&41&  21.&47&  4\\
3889    &H\,8+\ion{He}{1}
                       &0.&218&   17.&53&  20.&04&  5\\
3967    &[\ion{Ne}{3}] &0.&201&   21.&89&  22.&68&  4\\
4026    &\ion{He}{1}   &0.&193&    2.&02&   2.&11& 15\\
4067    &[\ion{S}{2}]  &0.&182&    1.&92&   0.&76& 23\\
4102    &H$\delta$     &0.&176&   24.&47&  26.&53&  4\\
4340    &H$\gamma$     &0.&128&   45.&48&  48.&77&  3\\
4363    &[\ion{O}{3}]  &0.&122&    3.&43&   3.&62& 10\\
4471    &\ion{He}{1}   &0.&094&    3.&97&   4.&27&  9\\
4563    &\ion{Mg}{1}]  &0.&070&    0.&69&   0.&71& 22\\
4571    &\ion{Mg}{1}]  &0.&068&    0.&31&   0.&32& 33\\
4861    &H$\beta$      &0.&000&  100.&00& 100.&00&  2\\
4881    &[\ion{Fe}{3}] &-0.&004&   0.&82&   0.&82& 21\\
4922    &\ion{He}{2}   &-0.&013&   1.&21&   1.&28& 17\\
4959    &[\ion{O}{3}]  &-0.&021& 100.&79&  99.&07&  2\\
5007    &[\ion{O}{3}]  &-0.&032& 303.&60& 296.&83&  1\\
5016    &\ion{He}{1}   &-0.&034&   2.&28&   2.&31& 12\\
5876    &\ion{He}{1}   &-0.&216&  11.&50&  10.&39&  6\\
6300    &[\ion{O}{1}]  &-0.&285&   2.&31&   2.&00& 12\\
6312    &[\ion{S}{3}]  &-0.&286&   2.&20&   1.&90& 13\\
6548    &[\ion{N}{2}]  &-0.&320&   5.&32&   4.&53&  8\\
6563    &H$\alpha$     &-0.&322& 359.&21& 305.&67&  1\\
6583    &[\ion{N}{2}]  &-0.&324&  15.&00&  12.&73&  5\\
6678    &\ion{He}{1}   &-0.&337&   3.&81&   3.&26& 10\\
6716    &[\ion{S}{2}]  &-0.&342&  27.&65&  23.&26&  4\\
6731    &[\ion{S}{2}]  &-0.&343&  19.&15&  16.&10&  4\\
7065    &\ion{He}{1}   &-0.&383&   2.&65&   2.&22& 11\\
7136    &[\ion{Ar}{3}] &-0.&391&   9.&53&   7.&83&  6\\
7281    &\ion{He}{1}   &-0.&406&   1.&37&   1.&14& 16\\
7319    &[\ion{O}{2}]  &-0.&410&   4.&75&   3.&87&  9\\
7330    &[\ion{O}{2}]  &-0.&411&   3.&85&   3.&13& 10\\
7751    &[\ion{Ar}{3}] &-0.&452&   2.&11&   1.&68& 13\\
\hline
\multicolumn{2}{l}{$EW$(H$\beta$), in \AA} &&& \multicolumn{5}{c}{285}\\
\multicolumn{2}{l}{$EW_{abs}$(H$\beta$), in \AA} &&& \multicolumn{5}{c}{2.0$\pm$0.1}\\
\multicolumn{2}{l}{$C$(H$\beta$), in dex} &&& \multicolumn{5}{c}{0.21$\pm$0.03}\\
\multicolumn{3}{l}{$F$(H$\beta$), in erg s$^{-1}$ cm$^{-2}$} &&\multicolumn{5}{c}{3.97$\times10^{-14}$}\\
\multicolumn{3}{l}{$I(H\beta)$, in erg s$^{-1}$ cm$^{-2}$} &&\multicolumn{5}{c}{6.21$\times10^{-14}$}\\
\enddata
\end{deluxetable}
\clearpage

\begin{deluxetable}{lccccc}
\tablecaption{Atomic Data for CELs in IRAF v2.15.1
\label{atomic}}
\tablewidth{0pt}
\tablehead{
Ion & Transition Probabilities & Collisional Strengths }
\startdata
N$^{0}$   & \citet{wie96} & \citet{dop76}  \\
N$^{+}$   & \citet{wie96} & \citet{len94}  \\
O$^{0}$   & \citet{wie96} & \citet{bha95}  \\
O$^{+}$   & \citet{wie96} & \citet{mcl93}  \\
O$^{++}$  & \citet{wie96} & \citet{len94}  \\
Ne$^{++}$ & \citet{kau86} & \citet{but94}  \\
S$^{+}$   & \citet{kee93} & \citet{ram96}  \\
S$^{++}$  & \citet{kau86} & \citet{gal95}  \\
Cl$^{++}$ & \citet{kau86} & \citet{but89}  \\
Ar$^{++}$ & \citet{kau86} & \citet{gal95}  \\
Ar$^{+3}$ & \citet{kau86} & \citet{zei87}  \\
\enddata
\end{deluxetable}
\clearpage

\begin{deluxetable}{lr@{}lr@{}lr@{}lcr@{}l}
\tablecaption{Densities and Temperatures
\label{tdat}}
\tablewidth{0pt}
\tablehead{
 &&&& NGC 456 &&&& \multicolumn{2}{c}{NGC 460} \\
\cline{2-7}
\cline{9-10}
 & \multicolumn{2}{c}{Position 1} & \multicolumn{2}{c}{Position 2} & \multicolumn{2}{c}{Position 3}&& \multicolumn{2}{c}{Position 1}}
\startdata
Densities (cm$^{-3}$)\\
$[$\ion{O}{2}$]$ &              $80$&$\pm15$ &             $130$&$\pm30$ &         $60$&$\pm30$        &&$170$&$\pm20$ \\
$[$\ion{S}{2}$]$ & \multicolumn{2}{c}{$<250$}&             $250$&$\pm60$ &  \multicolumn{2}{c}{$<500$} &&\multicolumn{2}{c}{$<150$}\\
$[$\ion{Cl}{3}$]$&\multicolumn{2}{c}{$<3000$}&\multicolumn{2}{c}{$<1000$}&\multicolumn{2}{c}{$<1300$}  &&  \mcnd       \\
\\
\hline
\\
Temperatures (K)\\ 
$[$\ion{O}{2}$]$ &     $13500$&$\pm250$& $12400$&$\pm400$ & $12600$&$\pm800$ && $12600$&$\pm600$ \\
$[$\ion{N}{2}$]$ &     $11300$&$\pm850$& $11700$&$\pm1000$&            \mcnd &&          \mcnd   \\
$[$\ion{S}{2}$]$ &      $8500$&$\pm700$&  $9700$&$\pm450$ & $7300$&$\pm1400$ &&  $8500$&$\pm450$  \\
$[$\ion{O}{3}$]$ &     $12650$&$\pm200$& $12165$&$\pm160$ & $12300$&$\pm550$ && $12400$&$\pm450$  \\
\enddata
\end{deluxetable}
\clearpage

\begin{deluxetable}{lccccccccccc}
\tabletypesize{\scriptsize}
\tablecaption{Ionic Abundance Determinations from Recombination
Lines\tablenotemark{a}
\label{trionic}}
\tablewidth{0pt}
\tablehead{
\colhead{Ion}
&\multicolumn{8}{c}{NGC 456} \\
& \multicolumn{2}{c}{Position 1} && \multicolumn{2}{c}{Position 2} &&\multicolumn{2}{c}{Position 3}\\
\cline{2-3}
\cline{5-6}
\cline{8-9}
&\colhead{$t^2 = 0.000$}&\colhead{$t^2 = 0.035\pm0.032$}
&&\colhead{$t^2 = 0.000$}&\colhead{$t^2 = 0.067\pm0.013$}
&&\colhead{$t^2 = 0.000$}&\colhead{$t^2 = 0.040\pm0.040$}
}
\startdata
He$^+$    & 10.910$\pm$0.011 & 10.906$\pm$0.012 && 10.923$\pm$0.008 & 10.918$\pm$0.008 && 10.945$\pm$0.015 & 10.924$\pm$0.019  \\
C$^{++}$  &  \nodata         &  \nodata         &&  7.47$\pm$0.18   &  7.46$\pm$0.18  &&  \nodata          &  \nodata          \\
O$^{++}$  &  \nodata         &  \nodata         &&  8.20$\pm$0.19   &  8.20$\pm$0.19  &&  \nodata          &  \nodata          \\
\\
\hline
\\
& \multicolumn{2}{c}{NGC 460 Position 1}  \\
& $t^2 = 0.000$    &  $t^2 = 0.036\pm0.027$\\
\cline{2-3}
He$^+$    & 10.935$\pm$0.014 & 10.921$\pm$0.015 \\
O$^{++}$  &  7.86$\pm$0.30   &  7.86$\pm$0.30   \\
\enddata
\tablenotetext{a} {In units of $12 +$ Log $N$(X$^{+i}$)/$N$(H),
 gaseous content only.}
\end{deluxetable}
\clearpage

\begin{deluxetable}{lccccccccccc}
\tabletypesize{\scriptsize}
\tablecaption{Ionic Abundance Determinations from Collisionally Excited 
Lines\tablenotemark{a}
\label{tceionic}}
\tablewidth{0pt}
\tablehead{
\colhead{Ion}
&\multicolumn{8}{c}{NGC 456} \\
&\multicolumn{2}{c}{Position 1}   && \multicolumn{2}{c}{Position 2}  &&  \multicolumn{2}{c}{Position 3} \\
\cline{2-3}  \cline{5-6}   \cline{8-9}
&\colhead{$t^2 = 0.000$}&\colhead{$t^2 = 0.035\pm0.032$}
&&\colhead{$t^2 = 0.000$}&\colhead{$t^2 = 0.067\pm0.013$}
&&\colhead{$t^2 = 0.000$}&\colhead{$t^2 = 0.040\pm0.040$} 
}
\startdata
C$^{++}$  &   \nodata     &   \nodata     && 6.75$\pm$0.15 & 7.28$\pm$0.18 &&  \nodata      &   \nodata     \\
N$^0$     &   \nodata     &   \nodata     && 4.54$\pm$0.08 & 4.71$\pm$0.06 &&   \nodata     &   \nodata     \\
N$^+$     & 6.03$\pm$0.03 & 6.15$\pm$0.14 && 5.92$\pm$0.04 & 6.07$\pm$0.04 && 5.90$\pm$0.07 & 6.02$\pm$0.16 \\
O$^0$     &   \nodata     &   \nodata     && 6.00$\pm$0.06 & 6.15$\pm$0.05 && 6.18$\pm$0.11 & 6.30$\pm$0.18 \\
O$^+$     & 7.31$\pm$0.03 & 7.46$\pm$0.18 && 7.35$\pm$0.05 & 7.53$\pm$0.04 && 7.38$\pm$0.10 & 7.53$\pm$0.19 \\
O$^{++}$  & 7.76$\pm$0.02 & 7.87$\pm$0.13 && 7.88$\pm$0.01 & 8.13$\pm$0.07 && 7.83$\pm$0.06 & 7.96$\pm$0.17 \\
Ne$^{++}$ & 6.96$\pm$0.03 & 7.09$\pm$0.14 && 7.11$\pm$0.02 & 7.39$\pm$0.07 && 7.10$\pm$0.07 & 7.25$\pm$0.19 \\
S$^+$     & 5.32$\pm$0.03 & 5.44$\pm$0.14 && 5.33$\pm$0.03 & 5.48$\pm$0.04 && 5.29$\pm$0.07 & 5.40$\pm$0.15 \\
S$^{++}$  & 6.24$\pm$0.05 & 6.37$\pm$0.14 && 6.30$\pm$0.03 & 6.58$\pm$0.07 && 6.03$\pm$0.12 & 6.18$\pm$0.22 \\
Cl$^{++}$ & 4.34$\pm$0.09 & 4.45$\pm$0.18 && 4.40$\pm$0.05 & 4.64$\pm$0.08 && 4.43$\pm$0.16 & 4.56$\pm$0.29 \\
Ar$^{++}$ & 5.58$\pm$0.03 & 5.68$\pm$0.11 && 5.64$\pm$0.02 & 5.86$\pm$0.06 && 5.42$\pm$0.06 & 5.54$\pm$0.16 \\
Ar$^{+3}$ &   \nodata     &   \nodata     && 3.88$\pm$0.12 & 4.14$\pm$0.14 &&  \nodata      &   \nodata     \\
\\
\hline
\\
& \multicolumn{2}{c}{NGC 460 Position 1}  \\
& $t^2 = 0.000$    &  $t^2 = 0.036\pm0.027$\\
\cline{2-3}
N$^+$     & 6.16$\pm$0.05 & 6.28$\pm$0.12 \\
O$^+$     & 7.57$\pm$0.06 & 7.73$\pm$0.16 \\
O$^{++}$  & 7.72$\pm$0.05 & 7.84$\pm$0.12 \\
Ne$^{++}$ & 7.03$\pm$0.06 & 7.16$\pm$0.14 \\
S$^+$     & 5.71$\pm$0.04 & 5.83$\pm$0.12 \\
S$^{++}$  & 6.30$\pm$0.08 & 6.43$\pm$0.14 \\
Ar$^{++}$ & 5.62$\pm$0.04 & 5.72$\pm$0.11 \\
\enddata
\tablenotetext{a} {In units of $12 +$ Log $N$(X$^{+i}$)/$N$(H),
 gaseous content only.}

\end{deluxetable}
\clearpage

\begin{deluxetable}{lccccccccc}
\tabletypesize{\small}
\tablecaption{NGC~456~Position~2 Gaseous Abundance Determinations\tablenotemark{a}
\label{tabund4}}
\tablewidth{0pt}
\tablehead{
\colhead{Element}             & 
\multicolumn{2}{c}{This paper}&& \colhead{PTP\tablenotemark{b}} &&
\colhead{PEFW\tablenotemark{c}} &&\colhead{GISFHP\tablenotemark{d}} \\
\cline{2-3}
\cline{5-5}
\cline{7-7}
\cline{9-9}
&\colhead{$t^2 = 0.000$}&\colhead{$t^2 = 0.067\pm0.013$}&&\colhead{$t^2 = 0.055$}&&\colhead{$t^2 = 0.000$}&&\colhead{$t^2 = 0.000$}
}
\startdata
He\tablenotemark{e} & 10.923$\pm$0.008 & 10.918$\pm$0.008 && 10.898  && \nodata &&  \nodata \\
C\tablenotemark{e}  &  7.57 $\pm$0.18  &  7.56 $\pm$0.18  && \nodata && \nodata &&  \nodata \\ 
C\tablenotemark{f}  &  6.85 $\pm$0.10  &  7.38 $\pm$0.18  && \nodata && \nodata &&  \nodata \\
N\tablenotemark{f}  &  6.67 $\pm$0.02  &  6.89 $\pm$0.03  &&  6.61   &&   6.48  && 6.43$\pm$0.02\\
O\tablenotemark{e}  &  8.29 $\pm$0.19  &  8.29 $\pm$0.19  && \nodata && \nodata &&  \nodata \\
O\tablenotemark{f}  &  7.99 $\pm$0.02  &  8.23 $\pm$0.05  &&  8.12   &&    8.07 && 8.06$\pm$0.01\\
Ne\tablenotemark{f} &  7.25 $\pm$0.02  &  7.51 $\pm$0.05  &&  7.22   &&    7.12 && 7.22$\pm$0.02\\
S\tablenotemark{f}  &  6.45 $\pm$0.04  &  6.72 $\pm$0.13  && \nodata && \nodata && 6.46$\pm$0.02\\
Cl\tablenotemark{f} &  4.53 $\pm$0.05  &  4.78 $\pm$0.11  && \nodata && \nodata && 4.66$\pm$0.04\\
Ar\tablenotemark{f} &  5.76 $\pm$0.04  &  5.98 $\pm$0.13  && \nodata && \nodata && 5.77$\pm$0.01\\
\enddata
\tablenotetext{a} {In units of $12 +$ Log $N$(X)/$N$(H).}
\tablenotetext{b} {\citet{pei76}.}
\tablenotetext{c} {\citet{pag78}.}
\tablenotetext{d} {\citet{gus11}.}
\tablenotetext{e} {Recombination lines.}
\tablenotetext{f} {Collisionally excited lines.}
\end{deluxetable}
\clearpage

\begin{deluxetable}{ccccccc}
\tabletypesize{\small}
\tablecaption{NGC~456: Positions~1 and 3 Gaseous Abundance Determinations\tablenotemark{a}
\label{tabund53}}
\tablewidth{0pt}
\tablehead{
\colhead{Element}
&\multicolumn{2}{c}{Position 1} && \multicolumn{2}{c}{Position 3} \\
\cline{2-3}
\cline{5-6}
& \colhead{$t^2 = 0.000$} & \colhead{$t^2 = 0.035\pm0.032$}
&& \colhead{$t^2 = 0.000$} & \colhead{$t^2 = 0.040\pm0.040$}
}
\startdata
He\tablenotemark{b}    &   10.910$\pm$0.011 & 10.906$\pm$0.012  &&  10.945$\pm$0.015 & 10.924$\pm$0.019  \\
N\tablenotemark{c}     &    6.70 $\pm$0.02  &  6.79 $\pm$0.07   &&   6.57 $\pm$0.05  &  6.68 $\pm$0.09   \\
O\tablenotemark{c}     &    7.89 $\pm$0.02  & 8.01 $\pm$0.10    &&   7.96 $\pm$0.05  &  8.10 $\pm$0.13   \\
Ne\tablenotemark{c}    &    7.13 $\pm$0.02  & 7.26 $\pm$0.09    &&   7.26 $\pm$0.05  &  7.41 $\pm$0.13   \\
S\tablenotemark{c}     &    6.39 $\pm$0.04  & 6.52 $\pm$0.17    &&   6.20 $\pm$0.07  &  6.35 $\pm$0.21   \\
Cl\tablenotemark{c}    &    4.39 $\pm$0.03  & 4.50 $\pm$0.08    &&   4.48 $\pm$0.05  &  4.61 $\pm$0.08   \\
Ar\tablenotemark{c}    &    5.71 $\pm$0.05  & 5.81 $\pm$0.16    &&   5.55 $\pm$0.06  &  5.67 $\pm$0.19   \\
\enddata
\tablenotetext{a} {In units of $12 +$ Log $N$(X)/$N$(H).}
\tablenotetext{b} {Recombination lines.}
\tablenotetext{c} {Collisionally excited lines.}
\end{deluxetable}
\clearpage

\begin{deluxetable}{cccccc}
\tabletypesize{\small}
\tablecaption{NGC~460 Position 1 Gaseous Abundance Determinations\tablenotemark{a}
\label{tabund8}}
\tablewidth{0pt}
\tablehead{
\colhead{Element}             & 
\multicolumn{2}{c}{This paper}&& \colhead{PEFW\tablenotemark{b}} \\
\cline{2-3}
\cline{5-5}
& \colhead{$t^2 = 0.000$} & \colhead{$t^2 = 0.036\pm0.027$} && \colhead{$t^2 = 0.000$}
}
\startdata
He\tablenotemark{c}    &   10.935$\pm$0.014 & 10.921$\pm$0.015 && \nodata \\ 
N\tablenotemark{d}     &    6.60 $\pm$0.04  &  6.70 $\pm$0.07  && \nodata \\
O\tablenotemark{c}     &    8.09 $\pm$0.30  &  8.10 $\pm$0.30  && \nodata \\
O\tablenotemark{d}     &    7.96 $\pm$0.04  &  8.09 $\pm$0.09  && 8.07 \\
Ne\tablenotemark{d}    &    7.30 $\pm$0.03  &  7.45 $\pm$0.08  && 7.36 \\
S\tablenotemark{d}     &    6.50 $\pm$0.06  &  6.63 $\pm$0.16  && \nodata \\
Ar\tablenotemark{d}    &    5.85 $\pm$0.08  &  5.95 $\pm$0.14  && \nodata \\
\enddata
\tablenotetext{a} {In units of $12 +$ Log $N$(X)/$N$(H).}
\tablenotetext{b} {\citet{pag78}.}
\tablenotetext{c} {Recombination lines.}
\tablenotetext{d} {Collisionally excited lines.}
\end{deluxetable}
\clearpage

\begin{deluxetable}{lr@{$\pm$}lr@{$\pm$}lr@{$\pm$}lr@{$\pm$}lr@{$\pm$}lr@{$\pm$}lr@{$\pm$}l}
\tabletypesize{\scriptsize}
\tablecaption{NGC~456, NGC~6822 V, NGC~346, 30~Doradus, the Orion Nebula, and Protosolar Total Abundances Relative to O\tablenotemark{a} \label{tta}}
\tablewidth{0pt}
\tablehead{
\colhead{Element}  &
\multicolumn{2}{c}{NGC~456\tablenotemark{b}} &
\multicolumn{2}{c}{NGC~6822 V\tablenotemark{c}} &
\multicolumn{2}{c}{NGC~346\tablenotemark{d}} &
\multicolumn{2}{c}{30~Doradus\tablenotemark{e}} &
\multicolumn{2}{c}{Orion\tablenotemark{f}} & 
\multicolumn{2}{c}{Sun\tablenotemark{g}}
}
\startdata
12 + log He/H &$10.918$&0.008 &$10.909$ & 0.011&$10.900$&0.003 &$10.928$&0.003 &$10.988$&0.003 &$10.98$&0.01\\
12 + log O/H  &$ 8.33$ &0.05  &$ 8.45 $ & 0.06 &$ 8.21$ &0.06  &$ 8.62$ &0.05  &$ 8.77$ &0.03  &$ 8.73$&0.05\\
log C/O	      &$-0.83$ &0.37  &$-0.34 $ & 0.13 &$-0.93$ &0.08  &$-0.48$ &0.05  &$-0.25$ &0.04  &$-0.26$&0.15\\
log N/O       &$-1.44$ &0.09  &$-1.40 $ & 0.17 &$-1.40$ &0.15  &$-1.27$ &0.08  &$-1.04$ &0.10  &$-0.86$&0.17\\
log Ne/O      &$-0.82$ &0.11  &$-0.82 $ & 0.09 &$-0.89$ &0.06  &$-0.79$ &0.06  &$-0.72$ &0.08  &$-0.76$&0.14\\
log S/O       &$-1.51$ &0.19  &$-1.65 $ & 0.09 &$-1.65$ &0.12  &$-1.63$ &0.10  &$-1.55$ &0.05  &$-1.57$&0.13\\
log Cl/O      &$-3.55$ &0.17  &$-3.74 $ & 0.10 &\mcnd          &$-3.70$ &0.12  &$-3.44$ &0.05  &  \mcnd     \\
log Ar/O      &$-2.35$ &0.18  &$-2.39 $ & 0.08 &$-2.39$ &0.10  &$-2.36$ &0.10  &$-2.15$ &0.06  &$-2.29$&0.13\\
log Fe/O      & \mcnd         &$-1.44 $ & 0.10 &$-1.47$ &0.10  &\mcnd          &$-1.27$ &0.20  &$-1.19$&0.11\\
\enddata
\tablenotetext{a}{The O and C gaseous abundances have been corrected for the fractions of these 
elements trapped in dust grains, see text.}
\tablenotetext{b}{Values in this column are relative to O with $t^2=0.067\pm0.013$.}
\tablenotetext{c}{Nebular abundances, values for $t^2 = 0.076 \pm 0.018$,
obtained for NGC 6822,from \citep{pea05}, with the exception of the Fe/O value that
comes from stellar data \citep{ven01}.}
\tablenotetext{d}{\citet{duf82,pei00,rel02}; \citet*{pea02}, values
for $t^2$ = 0.022$\pm$0.008. The Fe/O value comes from stellar data \citep{ven99,rol03,hun05}.}
\tablenotetext{e}{\citet{pea03}, values for $t^2$ = 0.033$\pm$0.005.}
\tablenotetext{f}{\citet{est04}, values for $t^2$= 0.022$\pm$0.002. The O and C abundances have been increased by 0.08 dex and
0.10 dex respectively to take into account the fractions of these elements trapped in dust grains. The Cl abundance has been decreased by 0.13 dex due to an error of +1.00 dex in the determination of the Cl$^+$/H$^+$ ratio.}
\tablenotetext{g}{Taken from the protosolar abundances of \citet{asp09}.}

\end{deluxetable}
\clearpage

\begin{deluxetable}{lr@{$\pm$}lr@{$\pm$}lcr@{$\pm$}lr@{$\pm$}l}
\tabletypesize{\scriptsize}
\tablecaption{NGC~456 and NGC~460 Total Abundances Relative to O\tablenotemark{a} \label{ttta}}
\tablewidth{0pt}
\tablehead{
\colhead{Element}  &
\multicolumn{4}{c}{NGC~456 Position 2} &&
\multicolumn{4}{c}{NGC~460 Position 1} \\
\cline{2-5}
\cline{7-10}
& \multicolumn{2}{c}{$t^2 = 0.000$} & \multicolumn{2}{c}{$t^2 = 0.067\pm0.013$}
&& \multicolumn{2}{c}{$t^2 = 0.000$} & \multicolumn{2}{c}{$t^2 = 0.036\pm0.027$}
}
\startdata
12 + log He/H            &$10.923$&0.008 &$10.918$&0.008 &&$10.935$&0.014 &$10.921$ & 0.015 \\
12 + log O/H             &$ 8.09$ &0.02  &$ 8.33$ &0.05  &&$ 8.06$ &0.04  &$ 8.19 $ & 0.09  \\
log C/O\tablenotemark{b} &$-0.82$ &0.37  &$-0.83$ &0.37  &&\mcnd          &\mcnd            \\
log C/O\tablenotemark{c} &$-1.24$ &0.12  &$-0.95$ &0.23  &&\mcnd          &\mcnd            \\
log N/O                  &$-1.42$ &0.03  &$-1.44$ &0.09  &&$-1.46$ &0.07  &$-1.49 $ & 0.17  \\
log Ne/O                 &$-0.83$ &0.03  &$-0.82$ &0.11  &&$-0.76$ &0.07  &$-0.74 $ & 0.17  \\
log S/O                  &$-1.64$ &0.06  &$-1.51$ &0.19  &&$-1.56$ &0.10  &$-1.56 $ & 0.26  \\
log Cl/O                 &$-3.56$ &0.06  &$-3.55$ &0.17  &&\mcnd          &\mcnd            \\
log Ar/O                 &$-2.32$ &0.06  &$-2.35$ &0.18  &&$-2.20$ &0.13  &$-2.24 $ & 0.23  \\
\enddata
\tablenotetext{a}{The O and C gaseous abundances have been corrected for the fractions of these
elements trapped in dust grains, see text.}
\tablenotetext{b}{Abundance from RLs relative to O from RLs.}
\tablenotetext{c}{Abundance from CELs relative to O from CELs.}
\end{deluxetable}
\clearpage

\clearpage

\end{document}